\documentclass[prl,twocolumn,showpacs,superscriptaddress]{revtex4}

\usepackage{graphicx}
\begin{document}

\title{Force network ensemble: a new approach to static
granular matter}

\author{Jacco H. Snoeijer} \affiliation{Instituut--Lorentz,
Universiteit Leiden, Postbus 9506, 2300 RA Leiden, The Netherlands}

\author{Thijs J. H. Vlugt} \affiliation{Department of Physical Chemistry
of Interfaces, Debye Institute, Utrecht University, Padualaan 8,
3584 CH Utrecht, The Netherlands}

\author{Martin van Hecke} \affiliation{Kamerlingh Onnes Lab,
Universiteit Leiden, Postbus 9504, 2300 RA Leiden, The Netherlands}

\author{Wim van Saarloos} \affiliation{Instituut--Lorentz,
Universiteit Leiden, Postbus 9506, 2300 RA Leiden, The Netherlands}

\date{\today}

\begin{abstract}
An ensemble approach for force distributions in static granular
packings is developed. This framework is based on the separation of
packing and force scales, together with an a-priori flat measure in
the force phase space under the constraints that the contact forces
are repulsive and balance on every particle. We show how the formalism
yields realistic results, both for disordered and regular triangular
``snooker ball'' configurations, and obtain a shear-induced unjamming
transition of the type proposed recently for athermal media.

\end{abstract}


\maketitle

The fascinating properties of static granular matter are closely
related to the organization of the interparticle contact forces into
highly heterogeneous force networks \cite{gm}. The probability density
of contact forces, $P(f)$, has emerged as a key characterization of a
wide range of thermal and athermal systems
\cite{network,edwards3D,qmodel,Pf,grest,liuletter,grestjam}. 
Most of these studies so far focussed on the broad tail of this
distribution. Recently, however, the non-universal shoulder of $P(f)$ for
small forces has received increasing attention, since it appears to
contain nontrivial information on the state of the system: $P(f)$
exhibits a peak at some small value of $f$ for ``jammed'' systems
which gives way to monotonic behavior above the glass transition
\cite{jamnote,liuletter,grestjam}. This hints at a possible connection
between jamming, glassy behavior and force network statistics, and
underscores the paramount importance of developing a theoretical
framework for the statistics and spatial organization of the forces.

A first step towards a statistical description of force networks is
the definition of a suitable ensemble over which to take averages. A
popular approach is that of Edwards, who proposed to take an equal
probability for all ``blocked'' states of a given energy and density
\cite{edwards,kurchan}. Each point in this ensemble defines a contact
geometry and a configuration of (balancing) contact forces. Even
though the precise particle locations and contact forces are related,
the crucial point is that for hard particles (most granular matter,
hard-sphere colloids, particles with a steep Lennard-Jones
interactions) a separation of scales occurs \cite{liuletter}: forces
inside a pile of steel balls originate from minute deformations
($\simeq 10^{-3}$).

\begin{figure}[t]
\includegraphics[width=8.0cm]{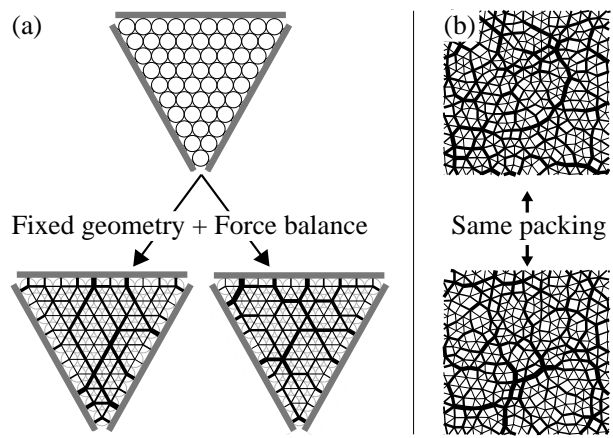} \centering
\caption{Two different mechanically stable force configurations for
{\em (a)} a ``snooker-triangle packing'' and for {\em (b)} an
irregular contact network of 1024 particles (only part is shown); the
thickness of the lines is proportional to the contact force.  The
``force network ensemble'' samples all possible force configurations
for a given contact network with an equal probability.}
\label{fig1}
\end{figure}

In this Letter we exploit this scale separation by focussing on
the {\em ensemble of force configurations for a given fixed
packing configuration} -- see Fig.~1. As we will show, the forces
can be considered underdetermined in this approach, since the
number of unknown forces exceeds the number of balance equations:
such packings are refered to as {\em hyperstatic}. 
Both packings of entirely rigid, frictional particles and packings 
of frictionless particles that deform $\simeq 10^{-3}$
are hyperstatic \cite{grestcoord}. For transparancy, however, we
will restrict ourselves to two dimensional frictionless
hyperstatic packings. In the spirit of Edwards, we sample all
allowed force configurations for a given packing with equal
probability. Interestingly, the idea to restrict the ensemble to
fixed geometries has also been suggested by Bouchaud in the
context of extremely weak tapping \cite{jp}.

Our ensemble approach captures the essential features of force
networks (Fig.~1) and leads to new insights on the non-universal
``shoulder'' of $P(f)$. In addition, by separating the contact
geometry from the forces, we can start to disentangle the separate
roles of contact and stress anisotropies in sheared systems.  The
conceptual advantage of not averaging over packing geometries is
complemented by practical advantages: our protocal is numerically cheap and
analytically accessible.

{\em Formulation} We study 2D packings of $N$ frictionless disks of
radii $R_i$ with centers ${\bf r}_i$. We denote the interparticle
force on particle $i$ due to its contact with particle $j$ by ${\bf
f}_{ij}$. There are $zN/2$ contact forces in such packings ($z$ being
the average contact number), and for purely repulsive central forces
we can write ${\bf f}_{ij} = f_{ij} {\bf r}_{ij}/|{\bf r}_{ij}|$,
where all $f_{ij} ~(=f_{ji})$ are positive scalars. For a fixed contact
geometry, we are thus left with $2N$ unknowns ${\bf r}_i$ and $z N/2$
unknowns $f_{ij}$ \cite{footcoornumber}.
These satisfy the conditions of mechanical equilibrium,
\begin{equation}
2N ~\mbox{eqs.:} \hspace*{3mm} \sum_j f_{ij}\,\frac{{\bf
r}_{ij}}{|{\bf r}_{ij}|} ={\bf 0}, \hspace*{5mm}\mbox{where } {\bf r}_{ij} = {\bf r}_i-{\bf r}_j,\label{equilibria}
\end{equation}
and once a force law $F$ is given, the forces are explicit functions of
the particle locations:
\begin{equation}
zN/2 ~\mbox{eqs.:} \hspace*{5mm} f_{ij} = F({\bf r}_{ij};R_i,R_j)~.
\label{feqs}
\end{equation}
Pakings of infinitely hard particles have $z=4$ and are thus {\em
isostatic}: For rigid particles Eqs.~(\ref{feqs}) reduce to $zN/2$
constraints on the $2N$ coordinates ${\bf r}_i$ which can only be
satisfied if $z\leq 4$, while Eqs.~(\ref{equilibria}) can only be
solved if $z \geq 4$; combining these yields $z=4$ \cite{isostatic,isostatic2}.

However, for particles of finite hardness, packings are typically {\em
hyperstatic} with $z>4$. 
A key parameter which quantifies the separation of length scales is
$\varepsilon = \frac{\langle f \rangle}{\langle r_{ij}\rangle} 
\langle\frac{dF_{ij}}{dr_{ij}}\rangle^{-1}$,
where $\langle \;\rangle$ denotes an average over the packing. 
We will avoid the strict isostatic $\varepsilon = 0$ case, but focus
instead on the regime where $\varepsilon$ is small and variations of
the force of order $\langle f \rangle$ result in minute variations of
${\bf r}_{ij}$, of relative size $\varepsilon$.  Hence, for
$\varepsilon \ll 1$, Eqs.~(1) and (2) can be considered separated, and the
essential physics is then given by the force balance constraints
Eqs.~(\ref{equilibria}) with fixed ${\bf r}_{i}$. In this
interpretation, there are more degrees of freedom $(zN/2)$ than
constraints $(2N)$, leading to an ensemble of force networks - see
Fig.~1.

It is important to note that different points in our ensemble do not
correspond to {\em precisely} the same packing of {\em exactly} the
same particles.  Our stochastic approach describes different force
configurations arising in, e.g., experiments on ``regular'' packings
of imperfect cannon balls \cite{duran} or packings under weak tapping
\cite{jp}. 
Experimentally, it has become clear that the macroscopic properties of
granular packings are sensitive to many (coarse grained) parameters
such as local densities, anisotropies and contact numbers, and that it
is very difficult to establish the relevant characteristics of a
packing. Our ensemble averages only over microscopic variations of the
packing, which have a strong effect on the local forces but not on
macroscopic properties, while keeping the important characteristics
fixed. 
The restriction to a minute part of the Edwards ensemble may 
therefore help to
disentangle the separate roles of contact and stress anisotropies.

The ensemble of force networks for a fixed contact geometry is
constructed as follows.  {\em{(i)}} Assume an a-priori flat
measure in the force phase space $\{f\}$. {\em{(ii)}} Impose the
$2N$ {\em linear} constraints given by the mechanical equilibrium
Eqs.~(\ref{equilibria}). {\em{(iii)}} Consider repulsive
forces only, i.e., $\forall f_{ij} \geq 0$.  {\em{(iv)}} Set an
overall force scale by applying a {\em fixed pressure}, similar to
energy or particle number constraints in the usual thermodynamic
ensembles.

We will first illustrate our formalism for a simple triangular
snooker-like packing (Fig.~\ref{fig1}a). Then, for an irregular
packing of 1024 particles (Fig.~\ref{fig1}b), we compare our ensemble
approach to MD simulations by varying the inverse ``hardness''
$\varepsilon$. The ensemble reproduces the $P(f)$ for sufficiently
hard particles well.  Finally, we find that applying a shear stress
yields an ``unjamming'' transition in our framework.

\begin{figure}[t]
\includegraphics[width=8.0cm]{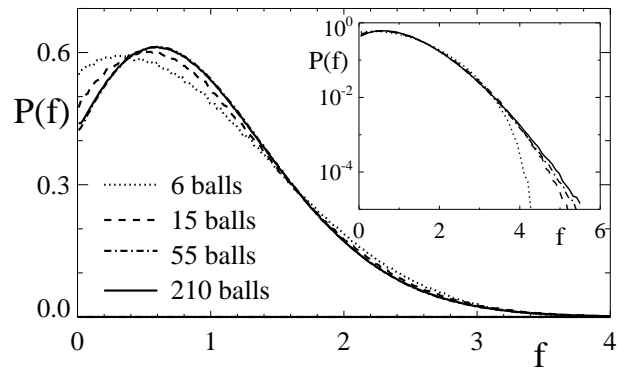} \centering
\caption{Interparticle force $P(f)$ for various triangular ``snooker''
packings (Fig.~1a). The inset shows the evolution of the tail for
large systems.}
\label{fig2}
\end{figure}

\paragraph{Regular packings}
The triangular snooker-like packings shown in Fig.~\ref{fig1}a have $3N$
unknown forces (boundary forces included) that are constrained by the
$2N$ equations of mechanical equilibrium.  Even though the packing
geometry is completely regular, the ensemble approach yields irregular
force networks and a broad $P(f)$.

Labeling each bond by a single index $k$, the mechanical equilibrium
can be expressed as
\begin{equation}\label{mecheq}
A\vec{f} = \vec{0} \quad\quad {\rm with} \quad \vec{f}\equiv
\left(f_1,f_2,\cdots,f_{3N} \right)~,
\end{equation}
where $A$ is a $3N$ times $2N$ sparse matrix. There is thus an
$N$-dimensional subspace of allowed force configurations that obey
mechanical equilibrium.  Eq.~(\ref{mecheq}) is homogeneous, but in
physical realizations an overall force scale is determined by the
externally applied stresses and/or the gravitational bulk forces. The
simplest manner to do so here is to fix the external pressure by
specifying the total boundary forces. For the snooker-triangles it
then follows that the sum of all forces is constant. We are thus
considering the phase space defined by the force balance
(\ref{mecheq}), the ``pressure'' constraint $\sum_k f_k = F_{\rm
tot}$, and the condition that all $f$'s are positive:
\begin{equation}\label{fullmatrixproblem}
{\cal A}\vec{f}= \vec{b} \quad\quad {\rm and} \quad \forall \quad
f_k\geq 0~, 
\end{equation}
where the fixed matrix ${\cal A}$ is the matrix $A$ extended by the
pressure constraint and $\vec{b}=(0,0,0,\cdots, 0, F_{\rm tot})$.

To compute $P(f)$ for larger packings, we have applied a simulated
annealing procedure \cite{numrec}. Starting from an ensemble of random
initial force configurations we sample the space of mechanically
stable networks, using a penalty function whose degenerate ground
states are solutions of Eq.~(\ref{fullmatrixproblem}). We have
carefully checked that results do not depend on the initial
configurations, and furthermore perfectly reproduce the distribution
$P(f)$ for 3 and 6 balls, which can be worked out analytically
\cite{longpaperagain}.

The two force networks shown in Fig.~1 are typical solutions $\vec{f}$
obtained by this scheme. 
We limit the discussion to $P(f)$ for interparticle forces, and
address the boundary forces which show different distributions
elsewhere \cite{rapid,longpaperagain}.  The interparticle $P(f)$ for
packings of increasing number of balls are presented in Fig.~2. Note
that all $P(f)$'s display a peak for small $f$, which is typical for
jammed systems \cite{liuletter}.  For large packings, this peak
rapidly converges to its asymptotic limit. The tail of $P(f)$ broadens
with system size, but the present data is not conclusive about its
asymptotic characteristics.

\begin{figure}[t]
\includegraphics[width=8.0cm]{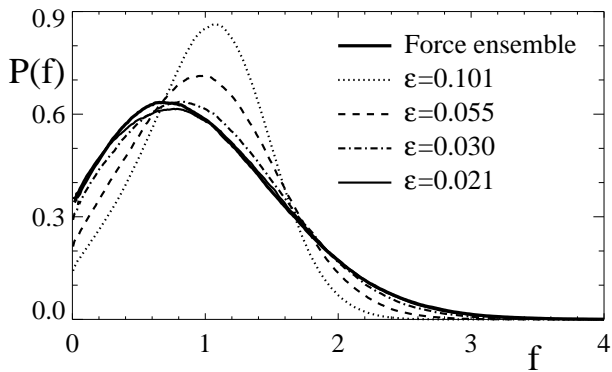} \centering \caption{Comparison
of the $P(f)$ obtained by our sampling of a frozen geometry, and MD
simulations of increasingly hard particles under constant pressure in
the limit $T \rightarrow 0$. }
\label{fig3}
\end{figure}

\paragraph{Irregular packings}
We now apply our force ensemble approach to a more realistic system
with a random packing geometry, and study a shear-stress induced
unjamming transition. To obtain a representative irregular contact
geometry, we perform a standard Molecular Dynamics simulation of a
50:50 binary mixture of 1024 particles with size ratio 1.4 that have a
purely repulsive 12-6 Lennard-Jones interaction (a shifted potential
with the attractive tail cut off), the same system as the one studied
by O'Hern {\em et al.} \cite{liuletter}.  We then quench such a finite
temperature simulation onto a $T=0$ random packing with a steepest
descent algorithm \cite{numrec}. The static contact network that we
obtain in this way then defines the matrix $\cal A$ in
Eq.~(\ref{mecheq}).

The $P(f)$ obtained for this fixed packing is displayed in
Fig.~\ref{fig3}: even for a single contact geometry, we clearly
reproduce a realistic $P(f)$ which is very similar to both that of the
triangular packings and to those obtained in experiments and
simulations \cite{network,Pf}.

To investigate the role of the particle hardness, we have performed MD
simulations of the same system with increasingly hard particles,
obtained by varying the prefactor of the potential at constant
pressure. For our original MD, which defined $\cal A$, the particles
are fairly soft with $\varepsilon \sim 0.1$, and the
corresponding $P(f)$ is somewhat different from the one in the force
ensemble. When the hardness of the particles is increased,
$\varepsilon$ diminishes and the corresponding $P(f)$ indeed
approaches the force ensemble $P(f)$. For the hardest particles
($\varepsilon \simeq 0.02$) these $P(f)$'s are virtually
indistinguishable. We find that this holds for a variety of force laws
\cite{longpaperagain}. This confirms the validity of our approach for
hard particles.

\paragraph{Unjamming by shear}
It is also possible to study the effect of a shear stress on the force
network ensemble by using the relation between the microscopic forces
and the macroscopic stress field:
\begin{equation}\label{stress}
\sigma_{\alpha\beta} = \frac{1}{V}\sum_k \, \left({\bf f}_{k}\right)_\alpha
\, \left({\bf r}_{k}\right)_\beta,
\end{equation}
and extending the matrix of Eq.~(\ref{fullmatrixproblem}) with the
three {\em linear} constraints of Eq.~(\ref{stress}). The average value of
the force is set to unity by requiring $\sigma_{xx}= \sigma_{yy}=1/2$
\cite{rfootnote}, and we vary $\tau=\sigma_{xy}/\sigma_{xx}$.

We find that $P(f)$ evolves from a ``jammed'' distribution with a
peak, to an ``unjammed'' monotonous distribution as a function of
shear stress (Fig.~4a). As a function of the angle $\phi$, $\langle\!f
\rangle\!$ varies in good approximation as $1 + 2 \tau \sin(2
\phi)$. This variation is consistent with Eq.~(\ref{stress}) as well
as with the alignment of the dominant contacts visible in Fig.~4b --
note the similarity to experimentally obtained sheared networks
\cite{shearclement}.  Since $P(f)$ contains forces in all directions,
the broadening of $P(f)$ with shear stress follows immediately from
this angular modulation of $\langle\!f \rangle\!$. However, this is
only part of the story: we find that the {\em shape} of $P(f)$ also
varies with direction, from extremely jammed along the force lines, to
almost purely exponential along the weak principle direction (not
shown) \cite{longpaperagain}.

When $\tau$ approaches $1/2$, $\langle\!f \rangle\!$ becomes zero along the
weak principle direction, which implies that all forces along
the weak direction approach zero. This can be  interpreted as
the breaking of contacts (see Fig.~4c). In addition, when
$\tau \rightarrow 1/2$, the contact number drops to $z=4$
(inset of Fig.~4a), and beyond this point, stable force
networks are no longer possible.
This simple mechanism thus provides $\tau=1/2$ as a definite upper
bound for the critical yield stress \cite{gm,nedderman,herrmann,duran}.
In terms of the ``slip angle'' this value corresponds to $30^\circ$.
On the other hand, it has been speculated
\cite{liuletter,grestjam} that the qualitative change of $P(f)$ from
peak to plateau could also be indicative of yielding; in our ensemble
this occurs at $\tau=0.26$ suggesting a slip angle of $15^{\circ}$.

\begin{figure}[t]
\includegraphics[width=8.0cm]{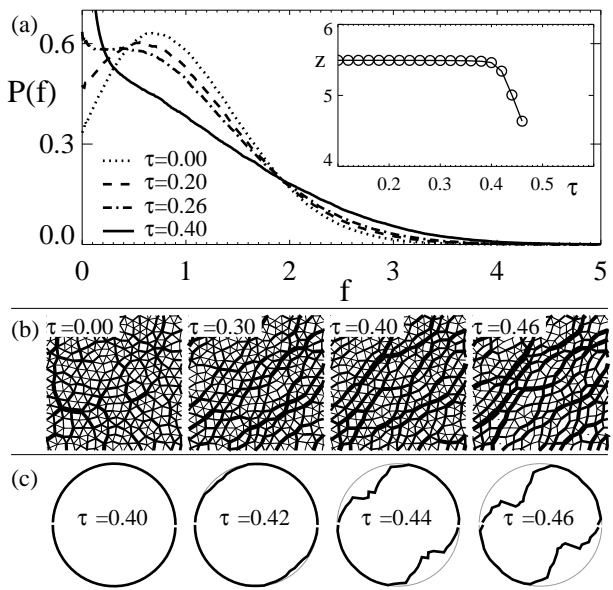} \centering \caption{Force
networks under shear. {\em (a)} $P(f)$ for increasing shear
showing an ``unjamming'' transition at $\tau \approx 0.26$. The
inset shows the contact number as function of shear, where
contacts are considered broken when $f< 10^{-4}$; for smaller
values of the cutoff this curve remains essentially the same. {\em
(b)} Examples of parts of the force networks under shear {\em (c)}
Ratio of number of contacts with $f>10^{-4}$ and number of
contacts as function of the contact angle show preferential
breaking along the ``weak'' principal direction. } \label{fig4}
\end{figure}

\paragraph{Outlook} We have proposes a novel ensemble approach to
athermal hard particle systems. The full set of mechanical equilibrium
constraints were incorporated, in contrast to more local
approximations or force chain models
\cite{qmodel,max_entropy,edwards3D,halseyinagroove,chainmodel}. A
number of crucial questions can possibly be addressed within our
framework. {\em{(1)}} Our approach is perfectly suited to include
frictional forces, since these are difficult to express in a force law
but simple to constrain by the Coulomb inequality. {\em{(2)}} The
contact and force networks of sand piles exhibit different
anisotropies under different construction histories
\cite{vanel,2Dbehringer}. We suggest that contact network anisotropies
may be sufficient to obtain the pressure dip under properly created
piles. {\em{(3)}} The problem defined by
Eqs.~(\ref{fullmatrixproblem}) could be generalized to arbitrary $\cal
A$ and $\vec{b}$, for which we can calculate $P(f)$ and may ask under
what conditions $P(f)$ has an exponential tail, appear jammed
etc. Preliminary work indicates that for realistic ${\cal A}$ but
taking $\vec{b}$ nonzero with mean square average proportional to $T$
captures the effect of a finite temperature of $P(f)$
\cite{longpaperagain}. 

We are grateful to Jan van der Eerden, Alexander Morozov and Hans van
Leeuwen for numerous illuminating discussions.  JHS gratefully
acknowledges support from the physics foundation FOM, and
MvH support from the science foundation NWO through a VIDI grant.


\begin{thebibliography}{99}

\small

\bibitem{gm} H.M. Jaeger, S.R. Nagel and R.P. Behringer,
Rev. Mod. Phys. {\bf 68}, 1259 (1996); P.G. de Gennes,
{\em ibid.} {\bf 71}, 374 (1999).

\bibitem{network} D.M. Mueth, H.M. Jaeger and S.R. Nagel,
Phys. Rev. E {\bf 57}, 3164 (1998);
D.L.  Blair {\em et al.}, {\em ibid.} {\bf 63}, 041304 (2001);
G. L\o voll, K.J. M\aa l\o y and E.G. Flekk\o y,
{\em ibid.} {\bf 60}, 5872 (1999).

\bibitem{edwards3D} J. Brujic {\em et al.},
Faraday Discussions {\bf 123}, 207 (2003).

\bibitem{qmodel} S.N. Coppersmith {\em et al.},
Phys. Rev. E {\bf 53}, 4673 (1996).

\bibitem{Pf} F. Radjai, M. Jean, J.J. Moreau and S. Roux,
Phys. Rev. Lett. {\bf 77}, 274 (1996);
S. Luding, Phys. Rev. E {\bf 55}, 4720 (1997);
F. Radjai, D.E. Wolf, M. Jean and J.J. Moreau,
Phys. Rev. Lett. {\bf 80}, 61 (1998);
A.V. Tkachenko and T.A. Witten, Phys. Rev. E {\bf 62}, 2510 (2000);
S.J. Antony, Phys. Rev. E {\bf 63}, 011302 (2000);
C.S. O'Hern, S.A. Langer, A.J. Liu and S.R. Nagel,
Phys. Rev. Lett. {\bf 88}, 075507 (2002).

\bibitem{grest} L. E. Silbert, G. S. Grest and J. W. Landry,
Phys. Rev. E {\bf 66}, 061303 (2002).

\bibitem{liuletter} C.S. O'Hern, S.A. Langer, A.J. Liu
and S.R. Nagel, Phys. Rev. Lett. {\bf 86}, 111 (2001).

\bibitem{grestjam} L.E. Silbert {\em et al.},
Phys. Rev. E {\bf 65}, 051307 (2002).

\bibitem{jamnote}A. J. Liu and S. R. Nagel, Nature {\bf 396}, 21
(1998); V. Trappe, V. Prasad, {\em et al.}, Nature {\bf 411}, 772
(2001).

\bibitem{edwards} S.F. Edwards and R.B.S. Oakeshott,
Physica A {\bf 157}, 1080 (1989).

\bibitem{kurchan} H.A. Makse and J. Kurchan,  Nature {\bf 415}, 614 (2002).

\bibitem{grestcoord} L.E. Silbert {\em et al.},
Phys. Rev. E {\bf 65}, 031304 (2002).

\bibitem{jp} J.P. Bouchaud, Proceedings of the 2002 Les Houches
Summer School on {\em Slow Relaxations and Nonequilibrium Dynamics in Condensed Matter}.

\bibitem{footcoornumber} Note that the number of unknown forces 
differs slightly from $zN/2$ if boundary forces are present.


\bibitem{isostatic} C.F. Moukarzel, Phys. Rev. Lett.
{\bf 81}, 1634 (1998).

\bibitem{isostatic2} A.V. Tkachenko and T.A. Witten,
Phys. Rev. E {\bf 60}, 687 (1999).

\bibitem{duran} J. Duran, {\em Sand, powders and grains},
Springer-Verlag, New York 2002.

\bibitem{numrec} W.H. Press {\em et al.} {\em Numerical Recipes: The
Art of scientific computing}, (Cambridge University Press, Cambridge,
England 1986).

\bibitem{longpaperagain} J.H. Snoeijer {\em et al.}, in preparation.

\bibitem{rapid} J.H. Snoeijer, M. van Hecke, E. Somfai and
W. van Saarloos, Phys. Rev. E {\bf 67}, 030302 (2003).

\bibitem{rfootnote} $|{\bf r}_k|$ is the average diameter of the two
particles in contact, and since we have verified that $f_k$ and $r_k$
do not correlate, we can take $r_k$ out of the sum.  Setting the trace
of $\sigma$ (i.e. the pressure) to unity then indeed leads to a
constraint of the form $\sum_k f_k = F_{\rm tot}$.


\bibitem{shearclement} J. Geng, G. Reydellet, E. Clement and
R. P. Behringer, Physica D {\bf 182}, 274 (2003).

\bibitem{nedderman} R.M. Nedderman {\em Statics and Kinematics of Granular
Materials}, (Cambridge University Press, Cambridge,
England 1992).

\bibitem{herrmann} J. Lee and H.J. Herrmann, J. Phys. A
{\bf 26}, 273 (1993);
R. Albert {\em et al.}, Phys. Rev. E {\bf 56}, R6271 (1997);
A. Daerr and S. Douady, Nature {\bf 399}, 241 (1999).

\bibitem{max_entropy} K. Bagi, Granular Matter {\bf 5}, 45 (2003).

\bibitem{halseyinagroove} T.C. Halsey and D. Ertas,
Phys. Rev. Lett. {\bf 83}, 5007 (1999).

\bibitem{chainmodel} J.P. Bouchaud, P. Claudin, D. Levine
and M. Otto, Eur. Phys. J. E {\bf 4}, 451 (2001).

\bibitem{vanel} L. Vanel {\em et al.}, Phys. Rev. E {\bf 60}, R5040 (1999).

\bibitem{2Dbehringer} J. Geng, E. Longhi, R.P. Behringer and
D.W. Howell, Phys. Rev. E {\bf 64}, 060301 (2001).


\end{thebibliography}
\end{document}